\documentclass[aps,prd,reprint,nofootinbib,groupedaddress,preprintnumbers,longbibliography]{revtex4-2}

\usepackage[T1]{fontenc} 
\usepackage{amsmath,mathrsfs}
\usepackage{graphicx}
\usepackage{subcaption}
\usepackage[]{hyperref}  
\usepackage[]{cleveref}
\usepackage{soul}
\usepackage{titlesec}

\usepackage{xcolor} 
\definecolor{c1}{HTML}{802410} 
\definecolor{c2}{HTML}{003262} 
\usepackage{tikz}
\usepackage{tkz-euclide}
\usetikzlibrary{decorations.pathmorphing}	
\tikzset{
    v/.style={decorate, decoration={snake, segment length=3mm, amplitude=0.75mm}, draw},
    f/.style={draw,decoration={markings,mark=at position #1 with {\arrow[very thick]{latex}}},postaction={decorate},node contents=#1},
    f/.default=.6,
    fb/.style={draw,decoration={markings,mark=at position #1 with {\arrowreversed[very thick]{latex}}},postaction={decorate},node contents=#1},
    fb/.default=.4,
    fnar/.style={draw},
    g/.style={decorate, draw,  decoration={coil,amplitude=3pt, segment length=3.5pt}},
    s/.style={dashed,draw, postaction={decorate},
        decoration={markings,mark=at position .55 with {\arrow[very thick]{latex}}}},
    sb/.style={dashed,draw, postaction={decorate},
        decoration={markings,mark=at position .55 with {\arrowreversed[draw=black,very thick]{latex}}}},
    snar/.style={dashed,draw,line width =1.25pt},
}
\usetikzlibrary{shapes}	
\tikzset{every picture/.style={line width=1}}

\newcommand{\Eq}[1]{Eq.~\hyperref[#1]{(\ref*{#1})}}
\newcommand{\Fig}[1]{Fig.~\hyperref[#1]{(\ref*{#1})}}

\newcommand{\sinc}{\operatorname{sinc}}
\usepackage{epstopdf}

\begin{document}

\title{Echoes of Global Cosmic Strings}

\author{Jeff Dror}
\email{jeffdror@ufl.edu}
\author{Antonios Kyriazis}
\email{akyriazis@ufl.edu}
\affiliation{Institute for Fundamental Theory, Physics Department, University of Florida, Gainesville, FL 32611, USA}

\date{\today}

\begin{abstract}
If the Universe underwent a cosmic phase transition, it may have left behind a network of cosmic strings. When these strings arise from the breaking of a gauge symmetry, their decay produces a significant stochastic background of gravitational waves. In contrast, if they originate from the breaking of a global symmetry, their decay predominantly yields Nambu–Goldstone bosons, which can persist as dark matter or dark radiation. In this work, we assess the detectability of this particle spectrum using a range of cosmological probes. We employ semi-numerical methods to estimate the resulting energy density and compute the associated matter power spectrum. We then compare these predictions with observations of the cosmic microwave background, Lyman-$\alpha$ forest, large-scale structure surveys, and the UV luminosity function, thereby deriving constraints on the Nambu–Goldstone boson mass and the symmetry-breaking scale. Finally, we present projections for the sensitivity of upcoming cosmic microwave background missions.
\end{abstract}

\maketitle

\section{\label{sec:level1} Introduction}
Whether the Universe underwent cosmological phase transitions remains one of the most profound open questions in cosmology. Such transitions can produce nearly topologically stable defects, including cosmic strings, depending on the underlying symmetry structure. If these defects later decay, their presence can be inferred through their decay products. Here, we focus on the global case, in which decay proceeds efficiently into (pseudo) Nambu--Goldstone bosons~\cite{Sikivie_2008,Gorghetto_2018,Gorghetto:2020qws,Buschmann_2022,Kim:2024wku,Dror_2021}. The detectability of these signals depends on the bosons’ mass and interactions. When their mass exceeds the Hubble scale, they can behave as a (possibly subdominant) component of dark matter. We therefore study the detectability of this spectrum through its minimal gravitational interactions, leveraging the extensive experimental program that measures the matter power spectrum, which is sensitive to perturbations sourced by the dark matter stress--energy tensor.

The gravitational influence of ultralight dark matter can be understood from the statistical properties of fluctuations in its stress--energy tensor~\cite{Hu:2000ke,Khmelnitsky:2013lxt,Hui:2016ltb}. Given a phase-space distribution, the induced metric perturbations can be described statistically~\cite{Kim:2023kyy,Kim:2024,Boddy:2025oxn,Dror:2025nvg}. Since the stress--energy tensor scales quadratically with the field, its temporal Fourier spectrum contains two characteristic classes of modes: (i) fast modes, with frequencies near twice the particle mass, corresponding to sums of single-particle energies; and (ii) slow modes, with frequencies set by the average kinetic energy, determined by the product of the mass and the velocity dispersion squared. Both classes fluctuate on the same characteristic spatial scale—the de~Broglie wavelength—but the slow modes dominate in amplitude and will be the focus of this study.

For a Maxwell--Boltzmann phase-space distribution, the statistical properties of these metric perturbations have been extensively studied. The associated density perturbations exhibit stochastic fluctuations on spatial separations larger than the coherence length, set by the inverse of the product of the particle mass and its velocity dispersion. This scale governs the structure of the matter power spectrum: for wavenumbers below the inverse coherence scale, the spectrum approaches a white-noise plateau, corresponding to a scale-independent (constant) isocurvature power spectrum. Together with the associated free-streaming length, this feature has motivated a rapidly developing program of cosmological searches for ultralight dark matter using probes of the matter power spectrum, including the $\textrm{Lyman-}\alpha$ forest, UV luminosity functions, large-scale structure surveys, and the cosmic microwave background (CMB)~\cite{Feix:2019lpo,Feix:2020txt,Irsic:2017yje,Kobayashi:2017jcf,Irsic:2019iff,Gorghetto:2021fsn,Gorghetto:2025uls,Chathirathas:2025aan,Gorghetto:2022ikz} (see also \cite{amin2024lowerbounddarkmatter,Harigaya:2025pox,Long:2024cak}). While these searches have rapidly become leading probes of ultralight dark matter from topological defect decay, they typically assume that the dominant constraints arise entirely from the white-noise plateau region of the matter power spectrum.

In this work, we develop a formalism to calculate the isocurvature power spectrum of ultralight dark matter for an arbitrary phase-space distribution, extending existing approaches that model the galactic dark matter field as a superposition of plane waves~\cite{Foster:2017hbq,Boddy:2025oxn} to a cosmological framework that consistently accounts for the expansion of the Universe. Using this formalism, we derive the matter power spectrum for ultralight dark matter produced via topological defect decay and characterize its behavior across all wavenumbers. We then estimate the sensitivity of existing and future cosmological probes. Furthermore, previous searches assumed a time-independent mass, such that the time of network collapse coincides with when the Hubble rate becomes comparable to its present-day mass; we go beyond this assumption by considering a temperature-dependent boson mass and studying its consequences.

To introduce the key results of our work, we present the form of the isocurvature power spectrum of post-inflationary pseudo Nambu--Goldstone bosons today $P_{\rm iso}(k)$, denoting the bosonic field as $\phi$ and assuming it forms a sub-component of dark matter:
\begin{equation}
    P_{\rm iso}(k) = D^{2}_{\rm iso}(k,a_{0})
    \left(\frac{\bar{\rho}_{\phi}(a_{0})}{\bar{\rho}_{\rm DM}(a_{0})}\right)^{2}
    k_\star^{-3}\,\mathcal{T}(k/k_\star).
    \label{eqn:general power spectrum}
\end{equation}
The power spectrum is proportional to the ratio of the relic density $\bar{\rho}_{\phi}$ to the total dark matter density $\bar{\rho}_{\rm DM}$, since fluctuations are defined relative to the total dark matter density. The growth factor $D_{\rm iso}(k,a_{0})$ accounts for the evolution of perturbations after matter--radiation equality. The transfer function $\mathcal{T}(x)$ tends to unity for small $x$, with a characteristic scale $k_\star$ above which it begins to fall. For cosmic string emission, the scale $k_\star$ is set by the infrared cutoff of the boson spectrum at the time the string network collapses, reflecting the fact that the emitted boson number density is dominated by particles with momenta near this cutoff scale. As a result, the features at $k>k_\star$ encode the information needed to distinguish between different ultralight dark matter production mechanisms. In this work, we calculate $\mathcal{T}(x)$ and determine $k_\star$, comparing the resulting isocurvature spectrum with that from adiabatic inflationary fluctuations.

The paper is organized as follows. In \Cref{sec:field correlations}, we construct the scalar field describing the bosons and derive its two-point correlation function. In \Cref{sec:density correlations}, we compute the density correlations of the slow modes in terms of the energy spectrum of particles emitted by cosmic strings, and derive the corresponding density power spectrum and transfer function. In \Cref{sec:constraints}, we impose constraints from large-scale structure data. We conclude in \Cref{sec:discussion} with a summary and a discussion of future directions.

\section{Field Correlations}
\label{sec:field correlations}
The evolution of a given mode of the scalar field $\phi$ in an expanding Universe is governed by the Klein--Gordon equation:~\footnote{We assume a quadratic potential, so the scalar field evolves linearly. Previous works have considered a cosine potential, which introduces non-linearities and leads to non-conservation of particle number after the collapse of the string network \cite{Gorghetto:2020qws,Chathirathas:2025aan,Gorghetto:2025uls,Gorghetto:2021fsn}.}
\begin{equation}
\ddot{\phi}_{\bf k}+ 3 H \dot{\phi}_{\bf k} + \phi_{\bf k}\left( m^{2} + \frac{|{\bf k}|^{2}}{a^{2}} \right) = 0,
\label{eqn:real k-g}
\end{equation}
where ${\bf k}$ is the comoving wavenumber, $a(t)$ is the scale factor, $H$ is the Hubble parameter, and overdots denote derivatives with respect to physical time.

The mass parameter may depend on temperature, as in the case of the QCD axion \cite{Sikivie_2008}. For temperatures $T > T_{\rm c} \equiv \sqrt{m_{a} f_{a}}$, the temperature dependence can be approximated as
\begin{equation}
    m(T) = m_{a} \left( \frac{T_{\rm c}}{T} \right)^{n},
\label{eqn:mass temperature}
\end{equation}
while for $T < T_{\rm c}$ one has $m(T) = m_{a}$, the zero-temperature boson mass~\cite{2018PhRvD..97h3502F,Maseizik:2024qly}. The temperature at which the field begins to oscillate, $T_m$, is determined numerically from $H(T_{m}) = m(T_{m})$, and the corresponding scale factor follows from entropy conservation. In this work, we consider both temperature-independent and temperature-dependent masses.

We model the ultralight dark matter field as a superposition of classical particles, treating each mode as a plane wave satisfying \Eq{eqn:real k-g}. Applying the Wentzel--Kramers--Brillouin (WKB) approximation, we obtain
\begin{align}
\phi_{\bf k}(x)
    &= \frac{\phi_{{\bf k},0}}{a^{3/2}}
       \sum_{j=1}^{N_{\bf k}}
       \cos\!\left[
           \int^{t}\!\omega_{\bf k}(t')\,dt'
           - {\bf k}\!\cdot\!{\bf x}
           + \varphi_{{\bf k},j}
       \right],
\label{eqn:label j}
\end{align}
where $x\equiv(t,{\bf x})$, $\omega_{\bf k}=\sqrt{m^{2}+{\bf k}^{2}/a^{2}}$, and $N_{\bf k}$ is the number of particles in a fixed phase-space volume (time-invariant since we work with comoving momentum). The sum over $j$ can be performed analytically (see, e.g., Refs.~\cite{Foster:2017hbq,Dror_2021,Boddy:2025oxn}), yielding
\begin{equation}
\phi_{\bf k}(x)
    = \sqrt{\frac{N_{\bf k}}{2a^{3}(t)}}\,\phi_{{\bf k},0}\,
      \alpha_{\bf k}
      \cos\!\left[
          \int^{t}\!\omega_{\bf k}(t')\,dt'
          - {\bf k}\!\cdot\!{\bf x}
          + \varphi_{\bf k}
      \right],
\label{eqn:phi with N}
\end{equation}
where $\alpha_{\bf k}$ and $\varphi_{\bf k}$ are random variables drawn from a Rayleigh distribution with unit scale parameter and a uniform distribution on $[0,2\pi)$, respectively~\cite{Foster:2017hbq}.

To fix the amplitude $\phi_{{\bf k},0}$, we compute the ensemble-averaged energy density of $\phi_{\bf k}$:
\begin{align}
\langle \rho_{\phi,{\bf k}}\rangle
    &= \left\langle
        \frac12\!\left(
            \dot{\phi}_{\bf k}^{2}
            + m^{2}\phi_{\bf k}^{2}
            + \frac{(\nabla\phi_{\bf k})^{2}}{a^{2}}
        \right)
      \right\rangle \notag\\
    &= \frac{N_{\bf k}}{2a^{3}}\,\phi_{{\bf k},0}^{2}\,\omega_{\bf k}^{2}.
\label{eqn:rho}
\end{align}
Equating this to the energy density of $N_{\bf k}$ particles, $\omega_{\bf k} N_{\bf k}/a^{3}$ per comoving volume element $d^{3}x$, gives
\begin{equation}
\phi_{{\bf k},0}
    = \sqrt{\frac{2}{d^{3}x\,\omega_{\bf k}(t)}}.
\label{eqn:phi0}
\end{equation}

The number of particles can be related to the phase-space density via $N_{\bf k}=f({\bf k})\,d^{3}x\,d^{3}k$. Because $f({\bf k})$ depends only on comoving momentum, it evolves self-similarly. Substituting \Eq{eqn:phi0} into \Eq{eqn:phi with N}, we obtain
\begin{align}
\phi_{\bf k}(x)
    &= \sqrt{
        \frac{f({\bf k})\,d^{3}k}
             {(2\pi)^{3}a^{3}\omega_{\bf k}}
       }\,
       \alpha_{\bf k}
       \cos\!\left[
           \int^{t}\!\omega_{\bf k}(t')\,dt'
           - {\bf k}\!\cdot\!{\bf x}
           + \varphi_{\bf k}
       \right].
\end{align}

\begin{figure}[t]
      \centering
    \includegraphics[width=1\columnwidth]{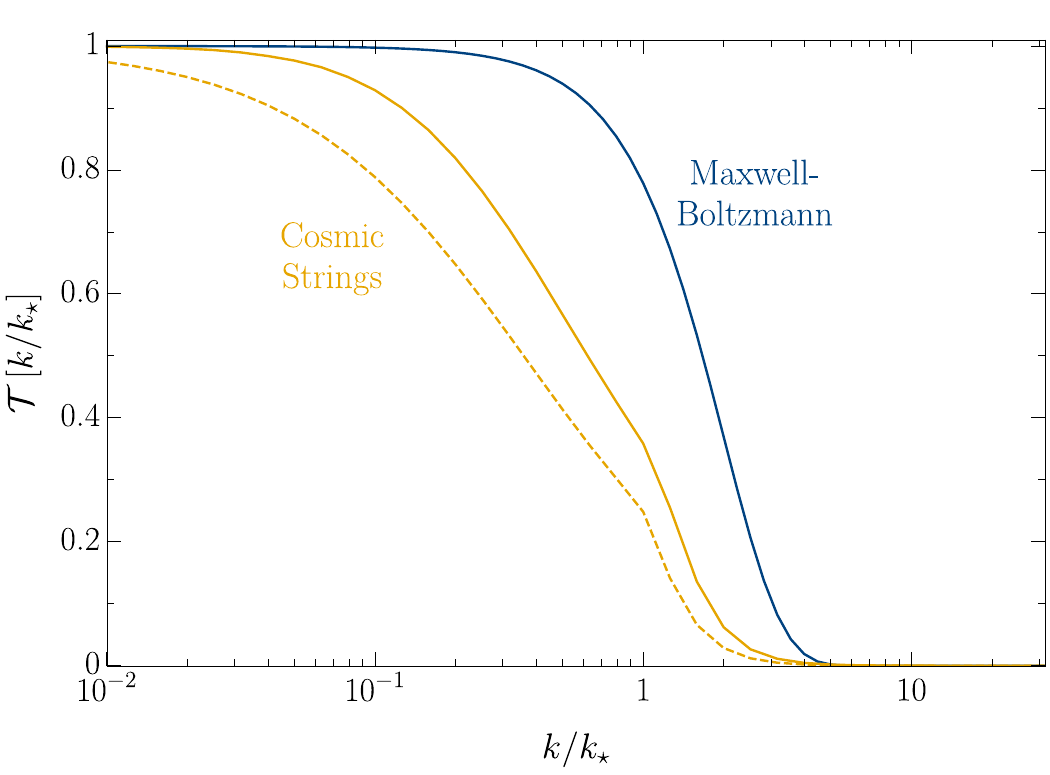}
    \caption{The Maxwell-Boltzmann (blue) and the cosmic strings (yellow) transfer functions. The solid line for cosmic strings corresponds to a numerical evaluation, while the dashed line corresponds to the analytic approximation presented in \Eq{eqn:transfer small k} and \Eq{eqn:transfer large k}, respectively. For $k \gg k_\star$, the cosmic string transfer function drops $ \propto k^{-4}$, while the Maxwell-Boltzmann transfer function drops exponentially. We have used $k_\star = 2.1 k_{\rm IR}$ to plot the cosmic string curves (see text).} 
    \label{fig:transfer functions}
\end{figure}

Using this expression, the full scalar field is constructed as a sum over plane-wave modes,
\begin{equation}
    \phi(x)=\sum_{\bf k}\phi_{\bf k}(x).
\end{equation}

To compute the two-point correlation function, we note that the ensemble average over the random phase $\varphi_{\bf k}$ vanishes for a single cosine, so only terms with ${\bf k}={\bf k}'$ contribute. We are primarily interested in the equal-time two-point function given by:
\begin{equation}
\langle \phi({\bf x},t)\phi({\bf x}',t)\rangle
    = \int\frac{d^{3}k}{(2\pi)^{3}}
      \frac{f({\bf k})}{a^{3}(t)\,\omega_{\bf k}(t)}
      \cos\!\big[{\bf k}\cdot\Delta{\bf x}\big],
\label{eqn:phi correl}
\end{equation}
where $\Delta{\bf x}={\bf x}-{\bf x}'$.

\section{The Matter-Power Spectrum}
\label{sec:density correlations}
\subsection{Density-Density Correlations}
The density is quadratic in $\phi$, which we treat as a Gaussian random field. Consequently, its two-point correlation function, $\langle \rho(x)\rho(x')\rangle$, can be reduced using Wick’s theorem to a sum of products of scalar two-point functions~\cite{Boddy:2025oxn}. This decomposition naturally separates contributions into ``fast'' modes, oscillating at frequencies $\sim 2m$, and ``slow'' modes, with frequencies set by the particle kinetic energies. In the non-relativistic limit, the fast-mode contribution is suppressed by $v^{2}$ relative to the slow component. Since we are interested in time-averaged observables, we retain only the slow-mode contribution in what follows.

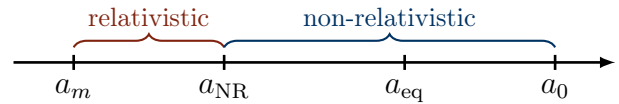
\begin{figure}[t!]
    \centering
    \begin{tikzpicture}
    \draw [-latex] (0,0)  coordinate (O) --++(8,0) coordinate (X) ;
    \foreach \x/\y/\z in {
    0.1/$a_m$/A,
    0.35/$a_{\rm NR}$/B,
    0.65/$a_{\rm eq}$/C,
    0.9/$a_0$/F
    }
    {\draw ($(O)!\x!(X)+(0.,-0.1)$) node[below] {\large \y} coordinate(\z) --++ (0,0.2);}
    \draw[c1,thick, decorate,decoration={brace,amplitude=5}] ($(A)+(0,0.3) $) --($(B)+(0,0.3) $) node[midway,yshift=0.4cm]{\normalsize relativistic}; 
    \draw[c2,thick, decorate,decoration={brace,amplitude=5}] ($(B)+(0,0.3) $) --($(O)!0.9!(X)+(0.,0.2) $) node[midway,yshift=0.4cm]{\normalsize non-relativistic}; 
    \end{tikzpicture}
    \caption{A sketch of the important instances in the problem. From $a_{m}$, the collapse of the string network, up to time $a_{\rm NR}$, the particles evolve relativistically and afterwards, they contribute to the dark matter density. These events take place before $a_{\rm eq}$, the time of radiation-matter equality.}
    \label{fig:timeline}
\end{figure}

A key timescale which sets the physics of ultralight dark matter density perturbations is when the bulk of the boson population begins to evolve non-relativistically, $a_{\rm NR}$. As explained shortly, the dark matter evolution after this point is well approximated by that of cold dark matter (CDM) and can be captured by the standard CDM growth factor $D(k,a)$. If the mass of the boson does not reach its constant value before radiation-matter equality, the growth factor will differ from that of CDM. We restrict our attention to the parameter space where this does not occur. As such, we now focus on evaluating the spectrum when the dominant component of the pseudo Nambu Goldstone bosons becomes non-relativistic, a condition given by the equation:
\begin{equation}
\label{eqn:tnr}
    m_{\rm NR} a_{\rm NR} = k_\star,
\end{equation}
where $ m _{ {\rm NR}} \equiv m ( a _{ {\rm NR}} ) $. In the case of a temperature-independent mass $m _{\rm NR} = m_a$. This relation is fixed by demanding that the characteristic physical momentum of the particles at time $a_{\rm NR}$ is equal to the particles' mass $m _{\rm NR}$. For later times, the physical momentum is always smaller than the mass and the particles evolve non-relativistically. The key time scales are summarized in \Fig{fig:timeline}.

We now evaluate the power spectrum at $a_{\rm NR}$. In the non-relativistic limit, we may set $\omega_{\bf k} \simeq  m _{\rm NR} $ and obtain: 
\begin{align}
\begin{split}
      \langle \rho_{\phi}(\textbf{x}) \rho_{\phi}(\textbf{x}') \rangle =  \frac{m^{2}_{\rm NR}}{a^{6}_{\rm NR} } \int \frac{d^{3}k d^{3}k'}{(2 \pi)^{6}}   f(\textbf{k}) f(\textbf{k}')  \cos \left( \Delta \textbf{x} \cdot \Delta \textbf{k}   \right),
\end{split}
\label{eqn: rhorho master}
\end{align}
where $\Delta {\bf k} \equiv {\bf k}' - {\bf k }$. A generalized expression of the two-point function, valid outside of the highly non-relativistic limit, is provided in \Cref{sec:density correl}. 

The angular integrals can be performed explicitly, assuming that the phase space density is isotropic, and the result is:
\begin{equation}
    \langle \rho_{\phi}(\textbf{x}) \rho_{\phi}(\textbf{x}') \rangle = \frac{m^{2}_{\rm NR}}{a^{6}_{\rm NR} \pi^{4}} \left[ \int_0^\infty d k k^{2} f(k) \sinc(k |\Delta \textbf{x}|) \right]^{2},
\label{eqn: rhorho strings}\end{equation}
where $\textrm{sinc}(x)  \equiv \textrm{sin}(x)/x$.

It is convenient to express $f(k)$ in terms of the critical energy density per unit log frequency: 
\begin{equation}
\label{eqn:Omega def}
    \Omega_\phi (k,a) \equiv \frac{1}{\rho_{c}(a)} \frac{d \rho_\phi}{d \log k},
\end{equation}
where the critical energy density $\rho_{c}(a) \equiv 3 M^{2}_{\rm pl} H^{2}(a)$ is defined as a function of scale factor and $M_{\textrm{pl}} \simeq 2.4 \times 10^{18}~ \textrm{GeV}$ is the reduced Planck mass. Using the evolution of the relativistic density $\rho_{\phi}$ for scale factors $a_{m} < a < a_{\rm NR}$:
\begin{equation}
\label{eqn:density rela}
    \rho_{\phi}(a) = \frac{1}{a^{4}(t)} \int \frac{dk k^{3}}{2 \pi^{2}} f(k),
\end{equation}
Using \cref{eqn:Omega def}, we can relate the phase space density $f(k)$ and the energy density per unit log momentum $\Omega_{\phi}(k)$. At the point of string network collapse (labeled by scale factor $a_m$):
\begin{align}
\begin{split}
    f(k) & =  \frac{2 \pi^{2}}{k^{4}} \rho_{c}(a_{m}) a^{4}_{m} \Omega_{\phi}(k,a_m)\,.
    \label{eqn:f and Omega}
\end{split}
\end{align}
Plugging \Eq{eqn:f and Omega} into \Eq{eqn: rhorho strings} and using that $\rho_c\propto a^{-4}$, we obtain: 
\begin{align}
\begin{split}
     &\langle \rho_{\phi}(\textbf{x}) \rho_{\phi}(\textbf{x}') \rangle \\ &\hspace{0.5cm}=  4 m^{2}_{\rm NR}a^{2}_{\rm NR} \rho^{2}_{c} (a_{\rm NR}) \bigg[\int \frac{d k}{k^{2}} \Omega_{\phi} (k) \sinc (k |\Delta \textbf{x}|) \bigg]^{2}.
\label{eqn:rho correl cosmic strings}
\end{split}
\end{align}

\begin{figure*}[]
\includegraphics[width=14cm]{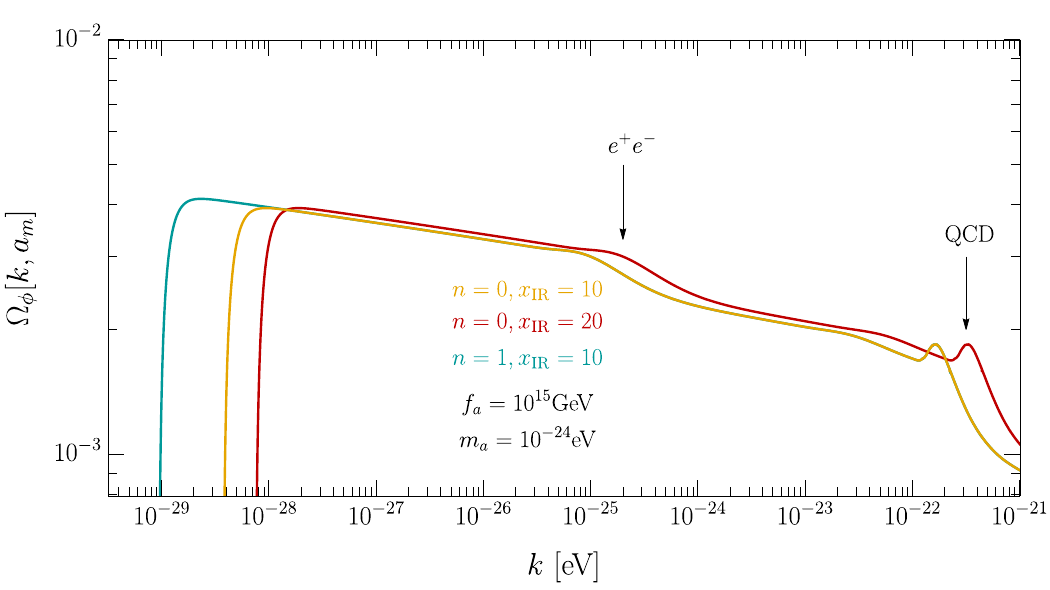}
    \caption{The spectrum $\Omega_{\phi}$ versus the frequency $k$, evaluated at $a_{m}$, for $m_{a}=10^{-24} \rm eV$, $f_{a} = 10^{15} \rm GeV$ and characteristic choices of $x_{\rm IR}$ and $n=0$. The yellow curve corresponds to $n=0, x_{\rm IR}=10$, the red curve to $n=0, x_{\rm IR} = 20$ and the cyan curve to $n=1, x_{\rm IR} = 10$. The $e^{+} e^{-}$ and $\rm QCD$ bumps are due to the changing relativistic degrees of freedom $g_{\ast}$ during the electron positron annihilation and the $\rm QCD$ phase transition respectively.}
    \label{fig:Omegak}
\end{figure*}

The spectrum of particles produced by cosmic string decay can be written as an integral over the string emission spectrum weighted by the string energy density. Because strings continuously radiate into Goldstone modes, their energy density approximately tracks the radiation energy density, leading to the so-called ``scaling solution.'' The emission spectrum is typically modeled as a power law, with a spectral index inferred from simulations (see Refs.~\cite{Sikivie_2008,Gorghetto_2018,Battye:2026whd,Dine:2020pds,Saikawa:2024bta} for reviews of the ongoing debate regarding this exponent). Assuming the string network follows a scaling solution, up to logarithmic corrections, allows the resulting particle spectrum to be evaluated numerically~\cite{Gorghetto_2018,Gorghetto:2020qws,Gorghetto:2022ikz,Gorghetto:2025uls,Benabou:2024msj,Battye:2026whd,Kaltschmidt:2025nkz,Saikawa:2024bta,Benabou:2023ghl,Hook:2026grn}. This formalism is reviewed in \cref{sec:scaling regime}.

An important characteristic of the spectrum is its infrared cut-off. This cut-off reflects the physical fact that cosmic strings longer than the horizon scale cannot efficiently radiate; instead, they reconnect and fragment into smaller loops. Consequently, the emitted radiation must be suppressed below a scale set by the inverse horizon size at a given time. Strings shorter than the horizon, on the other hand, decay by radiating Nambu--Goldstone bosons, producing a spectrum that peaks at
\begin{equation}
\label{eqn:kir}
    k_{\rm IR} = a_m H(a_m)\, x_{\rm IR},
\end{equation}
with $x_{\rm IR} \sim \mathcal{O}(10)$ \cite{Gorghetto_2018}. 

For temperature-independent masses, this cut-off is independent of $f_a$, while for $n\neq 0$ it shifts to smaller energies, as discussed in \cref{section:Cut-off}. The shape of the spectrum at larger energies depends on assumptions about the string decay mechanism. In what follows, we adopt the spectral index reported in \cite{Gorghetto_2018} and summarized in \cref{sec:scaling regime}, which yields a spectrum peaked at $k_{\rm IR}$ at late times.

We show the spectrum $\Omega_{\phi}$ in \Fig{fig:Omegak}, evaluated at $a_m$ for $m_a = 10^{-24}~\mathrm{eV}$ and $f_a = 10^{15}~\mathrm{GeV}$. For a temperature-independent mass, we display results for $x_{\rm IR} = 10$ (yellow) and $x_{\rm IR} = 20$ (red). We also show the case of a temperature-dependent mass with $n = 1$ and $x_{\rm IR} = 20$ (blue). In all cases, the infrared cut-off $k_{\rm IR}$ corresponds to the minimum value of $k$ for which $\Omega_\phi$ has support. The characteristic scale $k_\star$ denotes the peak of the spectrum, which appears at energies slightly above $k_{\rm IR}$ and is determined numerically in the following section. 

For fixed $x_{\rm IR}$, the spectra corresponding to different values of $n$ coincide at energies $k \gg k_{\rm IR}$. This behavior reflects the fact that high-energy modes are insensitive to the infrared cut-off, and in turn the temperature-dependent scaling of the mass. 
Increasing $x_{\rm IR}$ shifts the entire spectrum to larger energies; spectra with different $x_{\rm IR}$ are related by the rescaling $k \rightarrow (x_{\rm IR}/x_{\rm IR}') k$.

The features (``bumps'') visible in \Fig{fig:Omegak} arise from changes in the relativistic degrees of freedom $g_\ast$ during the expansion of the Universe. To identify the epoch corresponding to a given feature, we solve $k/(aH(a)) = x_{\rm IR}$ for $a$ at fixed $k$. Assuming radiation domination, this yields $a \propto T_{\rm eq}^2 a_{\rm eq}^2 x_{\rm IR}/(k M_{\rm pl})$. For $k \sim 10^{-22}~\mathrm{eV}$ and $k \sim10^{-25}~\mathrm{eV}$, as seen in \Fig{fig:Omegak}, we find that particles with these characteristic energies were first emitted around the time of the QCD phase transition and the epoch of $e^+ e^-$ annihilation, respectively.

\subsection{The Power Spectrum}
With the formalism to calculate $\Omega_{\phi}(k)$ established, we now compute the density correlations in \Eq{eqn:rho correl cosmic strings} and the associated power spectrum, defined through the Fourier transform:
\begin{equation}
    P_{\phi}(k,a_{\rm NR}) = \int d^{3}x \, e^{-i \mathbf{k} \cdot \mathbf{x}}
    \frac{\langle \rho_{\phi}(\mathbf{x}) \rho_{\phi}(0) \rangle}
    {\bar{\rho}_{\rm DM}(a_{\rm NR})^{2}} .
\end{equation}
This quantity is evaluated at the time when the particles become non-relativistic and begin contributing to the dark matter density. Performing the angular integral yields:
\begin{align}
\begin{split}
    & P_{\phi}(k,a_{\rm NR}) =  16 \pi m^{2}_{\rm NR} a^{2}_{\rm NR} \left( \frac{ a^{3}_{\rm NR} \rho_{c}(a_{\rm NR})}{\bar{\rho}_{\rm DM}(a_{0})} \right)^{2}  \\ &  \times \int^{\infty}_{0} dx x^{2} \sinc(k x) \left[ \int \frac{d k} {k^{2}} \Omega_{\phi} (k)\sinc \left( k x \right)  \right]^{2}. 
\end{split}
\label{eqn:power spec strings}\end{align}
Using the identity
\begin{align*}
   & \int^{\infty}_{0} d x x^{2} \sinc(k x) \sinc (\ell x) \sinc(p x)  \\ & \hspace{0.5cm}=\frac{\pi}{4 k p \ell}  
   \theta(p - |k - \ell|) \theta(k + \ell-p)\,,
\end{align*}
we may perform the remaining integral over the spatial separation $x$ leading to the power spectrum
\begin{align}
\begin{split}
     P_{\phi}(k,a_{\rm NR}) = 8 \pi^{2} \left( \frac{\bar{\rho}_{\phi}(a_{0})}{\bar{\rho}_{\rm DM}(a_{0})} \right)^{2} k_\star^{-3} \mathcal{T}\left( k \right),
\label{eqn: power specrum numerical}\end{split}
\end{align}
where the transfer function $\mathcal{T}(k)$ is
\begin{align}
\begin{split}
    \mathcal{T}\left( k \right) = & \frac{1}{2 k \mathcal{N}} \int^{\infty}_{k_{\rm IR}} \frac{dp}{p^{3}} \Omega_{\phi}(p)  \int^{k + p}_{|k - p|} \frac{d \ell}{\ell^{3}} \Omega_{\phi}(\ell)\,.
\end{split}
\label{eqn:transfer numerical}
\end{align}

Note that technically, we should not integrate over momenta for which the pseudo Nambu Goldstone bosons are relativistic. This imposes a condition on the integration limits set by the particle mass. Since the integrals are dominated by modes near $k_{\rm IR}$, and the relativistic cutoff lies parametrically above this scale for the masses considered here, we extend the upper limit to infinity with negligible error. The normalization of the transfer function $\mathcal{N}$ is chosen such that $\mathcal{T}(0) = 1$, fixing:
\begin{equation} 
    \mathcal{N} = \int^{\infty}_{k_{\rm IR}} \frac{dp}{p^{6}} \Omega_{\phi} (p)^{2},
\label{eqn:normalization}
\end{equation}
The relic energy density $\bar{\rho}_{\phi}$ is also determined as an integral over the spectrum:
\begin{align}
    \bar{\rho}_{\phi}(a_{0}) & = \int^{\infty}_{k_{\rm IR}} \frac{dk}{2 \pi^{2}} k^{2} m_{\rm NR} f\left(k\right)  \\ &  =  a^{4}_{\rm NR} \rho_{c}(a_{\rm NR}) m_{\rm NR}\int^{\infty}_{k_{\rm IR}} \frac{dk}{k^{2}} \Omega_{\phi}(k),
\label{eqn:relic density}
\end{align}
where we employ the convention that $a_{0}=1$. The characteristic momentum $k_\star$ is determined by the ratio between $\mathcal{N}$ and $(\bar{\rho}_{\phi}(a_{0})/a^{4}_{\rm NR} \rho_{c}(a_{\rm NR}) m_{\rm NR})^{2}$:
\begin{equation}
    k^{-3}_\star \equiv \left[ \int^{\infty}_{k_{\rm IR}} \frac{d \ell}{\ell^{2}} \Omega_{\phi}(\ell) \right]^{-2}\displaystyle \int^{\infty}_{k_{\rm IR}} \frac{dp}{p^{6}} \Omega^{2}_{\phi}(p)\,.
    \label{eq:kstar}
\end{equation} 
The characteristic momentum is largely set by the IR cutoff. In particular, we found numerically, that $ k_\star \simeq 2.1 k_{\rm IR}$ up to $1 \%$ accuracy for $10^{-27} ~\textrm{eV} \leq m_{a} \leq 10^{-18} ~\textrm{eV}$, $10^{14}~ \textrm{GeV} \leq f_{a} \leq 10^{15} ~\textrm{GeV}$ and $n=0,1$.

With these definitions, \Eq{eqn: power specrum numerical} is of the same form as \Eq{eqn:general power spectrum}, up to accounting for evolution between $a_{\rm NR}$ and $a_0$. The structure of the integrals in \Eq{eqn:transfer numerical} also matches the result of Ref.~\cite{Chathirathas:2025aan}.

In the literature, typically the dimensionless power spectrum is displayed $\Delta^{2}_{\phi}(k) = k^{3} P_{\phi}(k)/2 \pi^{2}$. Note that for $k \rightarrow 0$, we obtain from \Eq{eqn: power specrum numerical}:
\begin{equation}
    \Delta^{2}_{\phi}(k) = 4 \left(\frac{\bar{\rho}_{\phi}(a_{0})}{\bar{\rho}_{\rm DM}} \right)^{2} \left( \frac{k}{k_\star}\right)^{3}.
\label{eqn:dimensionless}\end{equation}
This matches with the results in Refs.~\cite{Gorghetto:2021fsn,Gorghetto:2025uls,Chathirathas:2025aan} up to a factor of 2 originating from our $k_\star$. 

To obtain the present-day spectrum we must also include the growth of perturbations after radiation–matter equality. This is encoded in the growth factor $D(k,a)$~\cite{Amin:2025ayf,Amin:2025sla,Gorghetto:2025uls}:~\footnote{Note that Ref.~\cite{Gorghetto:2025uls} calls this a transfer function, a name we reserve in this paper for $\mathcal{T}(k)$.}
\begin{equation}
\label{eqn:growth factor}
    D(k,a) \simeq \begin{cases}
        \displaystyle 1+\frac{a}{a_{\rm eq}} & k < k_{\rm J}(a_{\rm eq}) \\ \displaystyle \sqrt{ 1+ \frac{k_{\rm J}(a_{\rm eq})}{k} \left( \frac{a}{a_{\rm eq}} \right)^{2} } & k> k_{\rm J} (a_{\rm eq})
    \end{cases}
\end{equation}
The Jeans wavenumber is 
\begin{equation}
    \label{eqn:Jeans wavenumber}
    k_{\rm J}(a_{\rm eq}) = \sqrt{ \frac{a_{\rm eq}\rho_{\rm DM}(a_{0})}{2}} \frac{m_{a}}{M_{\rm pl}} \frac{1}{k_\star}\,.
\end{equation}
Comparing to $k_\star \simeq 2 k_{\rm IR}$,
\begin{align}
    \frac{k_{\rm J}(a_{\rm eq})}{k_\star} & = \left(\frac{1}{x_{\rm IR}} \right)^{2} \sqrt{\frac{\rho_{\rm DM}(a_{0})}{2 T^{4}_{\rm eq} a^{3}_{\rm eq}}} \left( \frac{M_{\rm pl}}{f_{a}} \right)^{\frac{n}{n+2}}\,, \\ 
    & \simeq 3 \times 10^{-3} \left( \frac{10}{x_{\rm IR}} \right)^{2} \left( \frac{M_{\rm pl}}{f_{a}} \right)^{\frac{n}{n+2}} \,.
\end{align}
Thus the Jeans scale is typically smaller than $k_\star$, mildly suppressing power for $k\lesssim k_\star$. A temperature-dependent mass brings the two scales closer, reducing this suppression.

\begin{figure*}[]
    \centering
    \includegraphics[width=14cm]{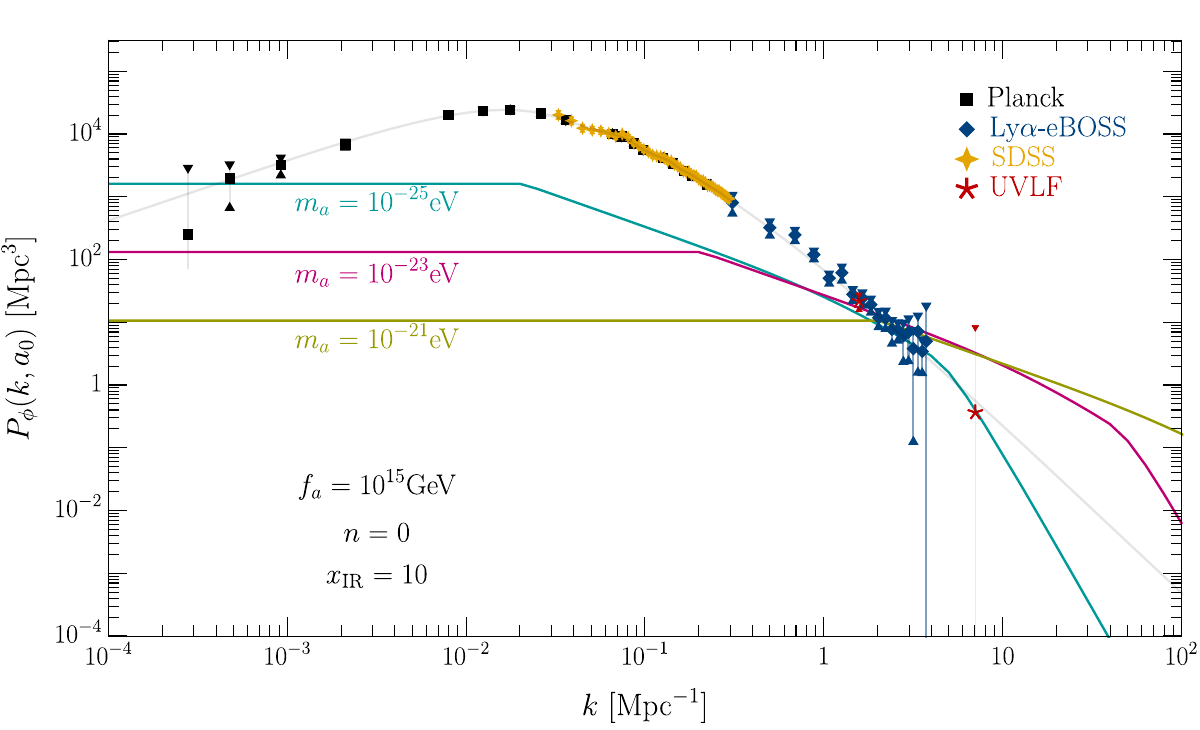}
    \caption{The power spectrum of pseudo-Goldstone bosons emitted by cosmic strings for $f_{a} = 10^{15} \rm GeV$, $n=0$ and $x_{\rm IR} = 10$. The three different curves correspond to $m_{a} = 10^{-25} \rm eV$ (cyan), $m_{a} = 10^{-23} \rm eV$ (purple) and $m_{a} = 10^{-21} \rm eV$ (olive green). The data of the observables were taken from $\rm Planck$ (black) \cite{Planck:2018vyg}, $\textrm{Lyman} \text{-} \alpha$ (blue) \cite{eBOSS:2018qyj},  UVLF (red) \cite{Sabti_2022}  and SDSS R7-(yellow) \cite{2010MNRAS.404...60R}. The triangles represent the 68\% confidence intervals for each measurement.}
    \label{fig:power spectrum strings}
\end{figure*}
    
The present-day spectrum is
\begin{equation}
\label{eqn:power spectrum today}
    P_{\phi}(k,a_{0}) = D(k,a_{0})^{2} P_{\phi}(k,a_{\rm NR})
\end{equation}
In \Fig{fig:power spectrum strings} we show this spectrum for $n=0$ and $x_{\rm IR}=10$ for three benchmark masses within observational reach. The discontinuity in the slope at small $k$ arises from $k_{\rm J}(a_{\rm eq})$, while a slope change at higher $k$ appears near $k\simeq k_\star$, as explained analytically in the next section. For $n\neq0$, the main difference is an overall enhancement of the spectrum at fixed $(m_a,f_a)$, since $k_\star$ decreases when $n\neq0$ (see \cref{section:Cut-off}). Previous work in this direction cut off the power spectrum at $k_\star$~\cite{Gorghetto:2021fsn,Gorghetto:2025uls}. The utility of our approach is to utilize the $k \geq k_\star$ components of the spectrum. 

\subsection{Analytic Approximation}
As is evident from \Fig{fig:Omegak}, $\Omega_{\phi}(k)$ varies slowly with $k$, which reflects the assumption that the string network reaches a near-scaling solution. In this section, we obtain analytic expressions for the power spectrum by approximating $\Omega_\phi$ as independent of $k$ for $k>k_{\rm IR}$ and zero for $k<k_{\rm IR}$.
The integrals in \Eq{eqn:normalization}, \Eq{eq:kstar}, and \Eq{eqn:relic density} can now be easily performed, leading to $\mathcal{N} \simeq \Omega^{2}_{\phi}/5$, $k_\star \simeq 5^{1/3} k_{\rm IR}$, and~\footnote{For $n=0$, this relic-density scaling agrees with the results of Refs.~\cite{Gorghetto:2021fsn,Gorghetto:2025uls} after using
$a^{4}_{\rm NR}\rho_{c}(a_{\rm NR})
= a^{4}_{m}\rho_{c}(a_{m})$
and the parametric relations
$\Omega_{\phi}\propto f_a^{2}$,
$a_m\propto m^{-1/2}$,
$k_\star\propto \sqrt{m}\,x_{\rm IR}$,
and
$\rho_c(a_m)\propto m^{2}$.}
\begin{align}
    \bar{\rho}_{\phi}(a_{0}) & \simeq
    \frac{a^{4}_{\rm NR} m_{\rm NR} \rho_{c}(a_{\rm NR}) \Omega_{\phi}}{k_\star},
\end{align}

Evaluating the transfer function integral in \Eq{eqn:transfer numerical} requires keeping track of the integration limits but is otherwise straightforward. For $k/k_{\rm IR} \leq 2 $, we can approximate its form as:
\begin{align}
\label{eqn:transfer small k}
       \mathcal{T} (y) & \simeq 
       \frac{5}{8y^{5} (1+y)}
      \bigg[ y ^5 + y^4 - 2y^3+6y^2 \\ &+12 y (1-\log (1+y))- 12  \log(1+y) \bigg] \notag
\end{align}
where $y = k/k_{\rm IR}$. When $k/k_{\rm IR} > 2 $,
\begin{equation}
       \mathcal{T} (y)  \simeq 
  \frac{-5}{4y^{5}} \left[ \frac{12y \hspace{-0.05cm}-\hspace{-0.05cm} 8y^{3}}{y^{2}-1} \hspace{-0.075cm}+\hspace{-0.075cm} 3 \log \left[ 
\frac{y+1}{y-1}  \frac{2y+1}{2y-1} \frac{2y^2 + y - 1}{2y^2 - y - 1} 
\right]\right] 
\label{eqn:transfer large k}
\end{equation}
The change in functional form at $y=2$ arises due to the absolute value at the lower limit of the inner integral in \Eq{eqn:transfer numerical} and requires treating the two regimes independently. The transfer function itself is valid for any value of $n$; the dependence on $n$ enters implicitly through the cutoff scale $k_{\rm IR}$. 

In \Fig{fig:transfer functions}, we compare this transfer function with the numerical result of \Eq{eqn:transfer numerical}. The two results diverge by at most $30 \%$ for $10^{-2} \lesssim k/k_\star \lesssim 1$. 

For $n=0$, an estimate for the order of magnitude of the power spectrum in the limit $k \rightarrow 0$ is then given by:
\begin{equation}
    \label{eqn:estimate P}
    P_{\phi} \simeq 800~\textrm{Mpc}^{3} \left( \frac{f_{a}}{10^{15} \textrm{GeV}} \right)^{4} \left( \frac{10^{-25} \textrm{eV}}{m_{a}} \right)^{1/2} \left( \frac{10}{x_{\rm IR}} \right)^{5}
\end{equation}
This matches the numerical results displayed in \Fig{fig:power spectrum strings} up to an $\mathcal{O}(1)$ factor. 

\section{Cosmic String Sensitivity}
\label{sec:constraints}
To search for a cosmic string signal, we construct a Gaussian likelihood from the measured power spectrum values, comparing the background-only likelihood $\mathcal{L}_{\rm B}$ to the signal-plus-background likelihood $\mathcal{L}_{{\rm S}+{\rm B}}$, which includes a contribution from the isocurvature power spectrum from \Eq{eqn: power specrum numerical}, $P_{\phi}(k)$. Explicitly, we have: 
\begin{equation}
    \mathcal{L}_{{\rm S}+{\rm B}} \propto \exp\left[- \sum_{i} \frac{(P_{\rm data}(k_{i}) - P_{\rm CDM}(k_{i}) - P_{\phi}(k_{i}))^{2}}{2 \sigma(k_{i})^{2}} \right],
\end{equation}
In this expression, $P_{\rm data}(k_i)$ denotes the observed matter power spectrum and $P_{\rm CDM}(k_i)$ the CDM prediction. The quantities $\sigma(k_i)$ are the experimental uncertainties. The background likelihood $\mathcal{L}_{\rm B}$ is obtained from $ {\cal L} _{ {\rm S} + {\rm B} } $ by setting $P_\phi(k)=0$.

For the CDM spectrum we adopt~\cite{Maggiore:2018sht}
\begin{equation}
    P_{\rm CDM}(k)
    = \frac{18\pi^2}{25}\,A_\Phi
    \left(\frac{a_{\rm eq}}{H_0^2\Omega_M}\right)^2
    k\,\mathcal{T}_{\rm CDM}^2(k)
    \left(\frac{k}{k_\ast}\right)^{n_s-1},
\label{eqn:CDM power spectrum}
\end{equation}
where $\mathcal{T}_{\rm CDM}(k)$ is the Bardeen--Bond--Kaiser--Szalay (BBKS) transfer function~\cite{Bardeen:1985tr}, $n_s=0.967$, $A_\Phi=0.95\times10^{-9}$, and $k_\ast=0.05~\mathrm{Mpc}^{-1}$~\cite{Planck:2018vyg}. In a more complete analysis, the CDM parameters would be varied jointly with $(m_a,f_a)$; however, previous studies indicate that the amplitude of the isocurvature spectrum is only weakly degenerate with these parameters~\cite{Gorghetto:2025uls}. Later, we estimate up to $\sim 30\%$ uncertainty in our limit on $ f _a $ when fixing them to their fiducial values.

Searching the matter power spectrum data for a signal, we do not find any significant evidence across any mass point. As such, we move to set constraints in the $(m_a,f_a)$ plane.  
The background consists of the adiabatic cold dark matter (CDM) perturbations, while the signal corresponds to the ultralight dark-matter--induced isocurvature perturbations derived in this work. We exclude a given parameter point $\{m_a,f_a\}$ when~\cite{2020NatRP...2..245A}
\begin{equation}
   2\log \left( \frac{\mathcal{L}_{{\rm S}+{\rm B}}}{\mathcal{L}_{\rm B}}\right) <-2.71,
\end{equation}
which corresponds to a one-sided $95\%$ confidence-level exclusion.

\begin{figure*}[]
    \centering
    \includegraphics[width=15cm]{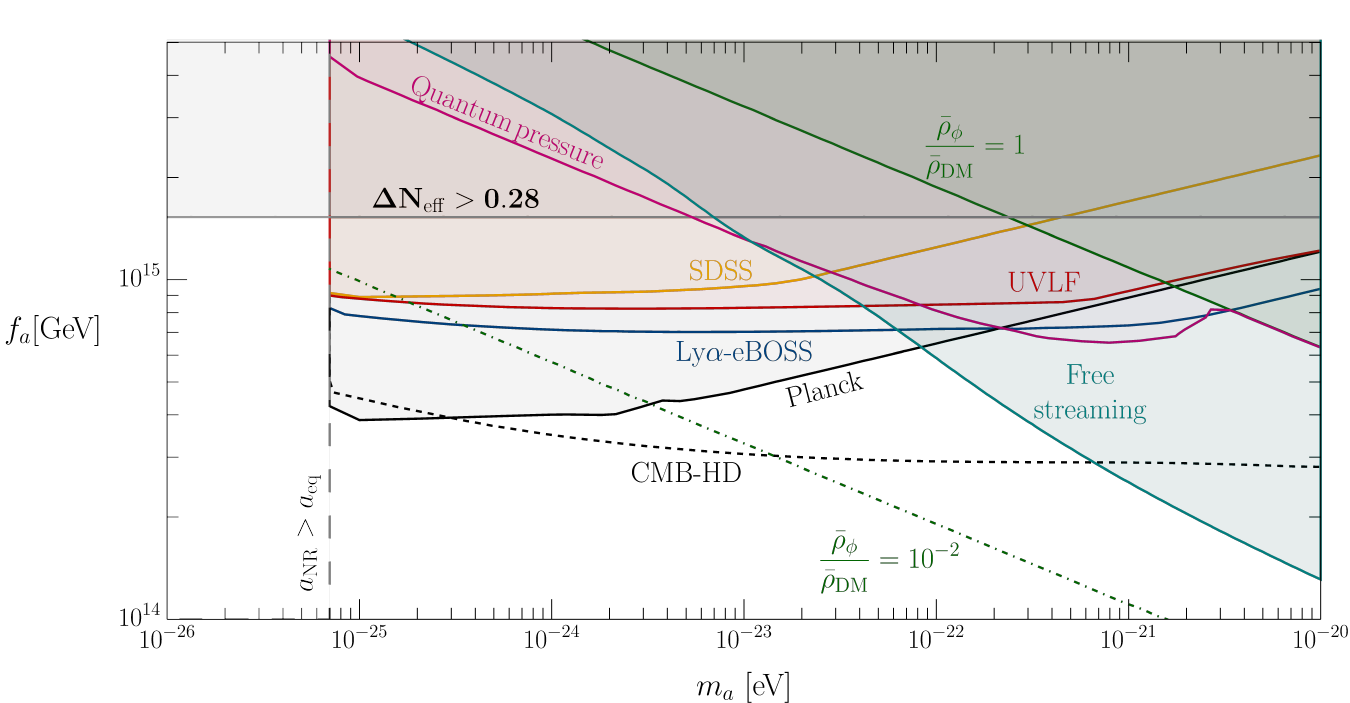}
    \caption{Constraints for cosmic strings in the parameter space of $m - f_{a}$ for $n=0$. The data of the observables were taken from $\rm Planck$ (black) \cite{Planck:2018vyg}, $\textrm{Lyman} \text{-} \alpha$ (blue) \cite{eBOSS:2018qyj},  UVLF (red) \cite{Sabti_2022}  and SDSS R7-(yellow) \cite{2010MNRAS.404...60R}. We also include the projections for a future $\rm CMB\text{-}HD$ mission \cite{macinnis2025cmbhdprobedarkmatter}. In pink, we depict the constraints from the quantum pressure \cite{Kobayashi:2017jcf} and in cyan the constraints from free-streaming, both affecting the adiabatic component of the power spectrum. The green region is where dark matter is overproduced. The $\Delta N_{\rm eff}$ bound is due to excessive radiation density during BBN \cite{Consiglio_2018,Kirilova:2023rnl}. Finally, for $a_{\rm NR} > a_{\rm eq}$, the bosons are relativistic after equality and our analysis does not apply.}
    \label{fig:constraints cosmic strings}
\end{figure*}

In \Fig{fig:constraints cosmic strings} we present the resulting constraints for $n=0$, using four matter power spectrum data sets: the Planck results~\cite{Planck:2018vyg}, Lyman$\text{-}$$\alpha$ forest measurements~\cite{eBOSS:2018qyj}, UV galaxy luminosity function data~\cite{Sabti_2022}, and SDSS DR7 measurements~\cite{2010MNRAS.404...60R}. 
We also show the projected sensitivity of a future CMB-HD lensing survey~\cite{macinnis2025cmbhdprobedarkmatter}. 
All data sets were compiled following Ref.~\cite{macinnis2025cmbhdprobedarkmatter}.

The constraints from the matter-power spectrum span a wider range of masses than those obtained in Refs.~\cite{Gorghetto:2021fsn,Gorghetto:2025uls} since these prior works focused on the influence of the white noise plateau on the matter power spectrum and assumed a zero value for smaller scales (larger wavenumbers). In particular, our constraints from the Planck data appear to be consistent with those presented in Fig. 4 of \cite{Buckley:2025zgh}. There, the isocurvature power spectrum was parameterized by an amplitude $A_{\rm iso}$ and a characteristic scale $k_\star$. Both were considered free parameters. The functional form of the power spectrum was considered as a broken power law, with the white noise plateau for $k < k_\star$ and a scaling $k^{-3}$ for $k>k_\star$. The constraints were depicted in the $\{ k_\star, A_{\rm iso} \}$ plane. They are particularly strong where $k_\star \leq 0.1 \textrm{Mpc}^{-1}$, as the CMB data are concentrated in this regime, while they scale as $k_\star^{-3}$ for larger wave-numbers. With the identifications $A_{\rm iso} \simeq \left( \bar{\rho}_{\phi} / \bar{\rho}_{\rm DM} \right)^{2}$ and $k_\star \simeq 3 \textrm{Mpc}^{-1} \left( m_{a} / 10^{-25} \textrm{eV} \right)^{1/2}$, we indeed find good agreement between the results presented in our \Fig{fig:constraints cosmic strings} and the results in \cite{Buckley:2025zgh}.
The flat regions of the constraints that are observed for the SDSS, Lyman\text{-}$\alpha$ and UVLF data occur because for these masses, the scales probed by these observables are within the white noise plateau of the isocurvature power spectrum. As we move to lower $m_{a}$, the characteristic scale is also reduced and it becomes increasingly important to consider the tail of the power spectrum for $k > k_\star$ in order to impose constraints from these observables. 
In \cite{Garcia-Gallego:2026phh,Kobayashi:2017jcf,Gorghetto:2025uls}, the Lyman$\text{-}$$\alpha$ constraints appear to be more stringent than the ones we present in this paper.That is because the underlying dataset that is used to impose these constraints originates from the MIKE/HIRES spectrometers \cite{Viel:2013fqw}, which yield higher resolution of individual sources and therefore a larger signal-to-noise ratio\cite{eBOSS:2018qyj}.These constraints are not relevant in the context of the present study, given that the white noise plateau was primarily used in those works, whereas we have attempted to include the high wavenumber tail in this analysis. Including the derivation of these constraints is distinct from the primary objectives of this paper and we defer this analysis to future work.

For large $m_{a}$, the power spectrum is reduced primarily because of the scaling $P_{\phi} \propto k^{-3}_\star \propto m^{-3/2}_{a}$ and the constraints are weakened. The kink that appears at $m_{a} = 4 \times 10^{-24} \rm eV$ in the curve of the Planck data is due to the discontinuity of our power spectrum at $k_{\rm J}(a_{\rm eq})$. This particular mass corresponds to $k_{\rm J}(a_{\rm eq}) \simeq 0.1~ \textrm{Mpc}^{-1}$, close to the maximum of the adiabatic power spectrum.

We also highlight that the increased sensitivity of a future $\textrm{CMB-HD}$ experiment can probe $f_{a}$ down to $f_{a} \simeq 4 \times10^{14} \textrm{GeV}$ for masses $ 3 \times 10^{-25} \textrm{eV} \lesssim m_{a} \lesssim 6 \times 10^{-22} \textrm{eV}$, well below the sensitivity of existing constraints.

The vertical dashed line at $m_{a} = 7 \times 10^{-26} \rm eV$ depicts the boundary beyond which the non-relativistic approximation that we have employed is not valid. Studying the bounds below this mass point requires treating the pseudo-Goldstone boson population relativistically as is beyond the scope of our analysis.

The upper portion of the plot is constrained by the contribution of relativistic Nambu--Goldstone bosons to the radiation density during Big Bang Nucleosynthesis, parameterized by $\Delta N_{\rm eff}$. Across the parameter space considered here, these particles remain relativistic at temperatures $T \sim \mathrm{MeV}$, so this bound is effectively independent of the mass~\cite{Consiglio_2018,Gorghetto:2021fsn}. The abundance of these relativistic particles, derived in \cref{sec:scaling regime}, depends only weakly on the onset of the scaling regime; in drawing this bound, we assume the scaling regime begins at a temperature $T \simeq f_a$.

For densities $\bar{\rho}_{\phi}(a_{0})/\bar{\rho}_{\rm DM}(a_{0}) \gtrsim
0.1$ the leading constraints come from quantum pressure and free-streaming. The wavenumber associated with quantum pressure is given by~\cite{Kobayashi:2017jcf,Hlozek:2014lca}:
\begin{equation}
    k_{\rm Q}(a_{\rm eq}) \simeq 7~\textrm{Mpc}^{-1} \left( \frac{m_{a}}{10^{-22} \textrm{eV}} \right)^{1/2}.
\end{equation}
The adiabatic spectrum is suppressed for wavenumbers larger than this one and this scale is relevant even if no velocity dispersion is assumed, in contrast to $k_{\rm J}$.  
The data to draw this bound were taken from \cite{Kobayashi:2017jcf}. 

The methodology to draw the free-streaming bound follows closely the work of Refs.~\cite{amin2024lowerbounddarkmatter,Long:2024cak,Liu:2024pjg} and is presented in \Cref{sec:free streaming}. Since particles with finite velocity dispersion are unable to cluster, they suppress the matter power spectrum below the co-moving free-streaming scale \cite{Liu:2024pjg,amin2024lowerbounddarkmatter}: 
\begin{equation}
    \lambda_{\rm fs}(k,a) =  \int^{a}_{0} \frac{da'}{(a')^{2} H(a')} \frac{k}{\sqrt{k^{2} + (a')^{2} m^{2}(a')}}
\label{eqn:free streaming scale}
\end{equation}
The free-streaming suppression is encoded in a transfer function that multiplies the adiabatic spectrum and is proportional to the dark matter density fraction of the sub-component \cite{amin2024lowerbounddarkmatter,Gorghetto:2025uls}. In the case where the energy spectrum $\Omega_{\phi}(k)$ is peaked at $k_\star$, an approximate form of this transfer function is:

\begin{equation}
\label{eqn:transfer free approx}
    \mathcal{T}_{\rm fs}(k,a_{0}) \simeq \frac{\bar{\rho}_{\phi}(a_{0})}{\bar{\rho}_{\rm DM}(a_{0})} \sinc \left( k \lambda_{\rm fs}(k_\star,a_{0}) \right).
\end{equation}

To draw the bound, we used the Planck-2018 data. In our approach, we used the numerical results for the spectrum $\Omega_{\phi}(k)$ and the full expression for the free-streaming transfer function provided in \cref{sec:free streaming} and confirmed that \Eq{eqn:transfer free approx} produces the same constraints. 

\begin{figure*}[]
    \centering
    \includegraphics[width=15cm]{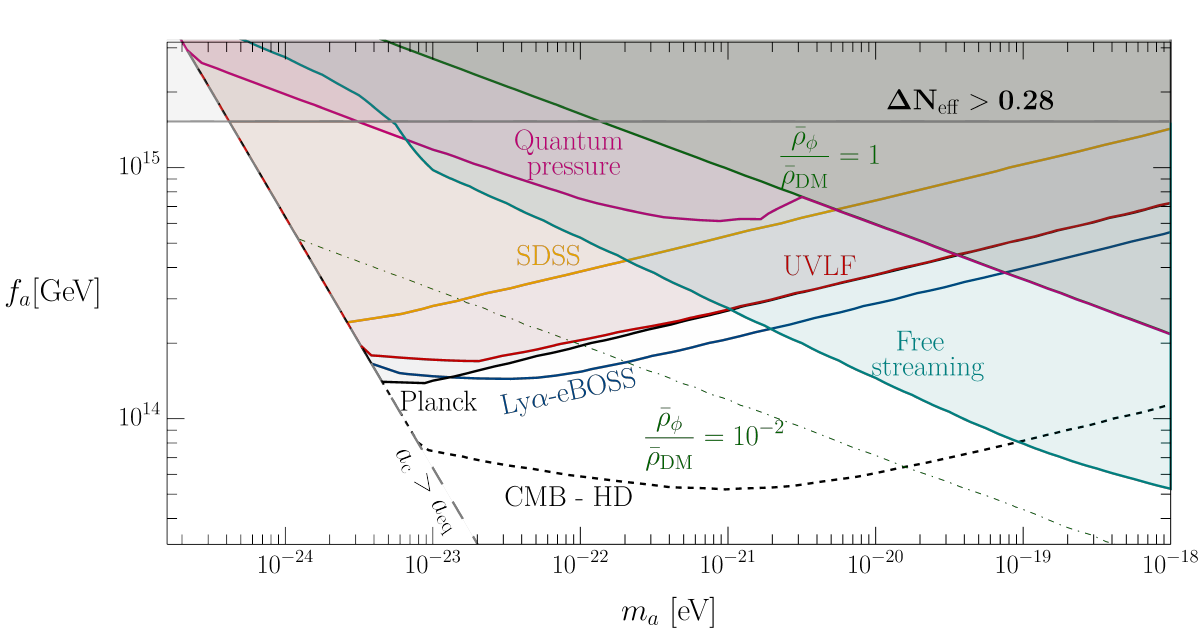}
    \caption{Constraints for cosmic strings in the parameter space of $m_{a} - f_{a}$ for $n=1$. The data of the observables were taken from $\rm Planck$ (black) \cite{Planck:2018vyg}, $\textrm{Lyman} \text{-} \alpha$ (blue) \cite{eBOSS:2018qyj},  UVLF (red) \cite{Sabti_2022}  and SDSS R7-(yellow) \cite{2010MNRAS.404...60R}. We also include the projections for a future $\rm CMB \text{-} HD$ mission \cite{macinnis2025cmbhdprobedarkmatter}.In pink, we depict the constraints from the quantum pressure \cite{Kobayashi:2017jcf} and in cyan the constraints from free-streaming, both affecting the adiabatic component of the power spectrum. The green region is where dark matter is overproduced. The $\Delta N_{\rm eff}$ bound is due to excessive radiation density during BBN \cite{Consiglio_2018,Kirilova:2023rnl}. For $  a_{\rm c} \equiv a (T_{\rm c})>a_{\rm eq}$ the mass obtains its zero temperature value after equality, where our analysis does not apply.}
    \label{fig:constraints cosmic strings n=1}
\end{figure*}

In \Fig{fig:constraints cosmic strings n=1}, we show the constraints discussed above for $n=1$. For $m_{a}$ and $f_{a}$ to the left of the line indicated by the label $a_{\rm c} \equiv a (T_{\rm c}) > a_{\rm eq} $, the mass reaches its constant value after equality. In this regime, the growth factor $D_{\rm iso}(k,a)$ is modified, and our analysis is not directly applicable. The region where the relativistic evolution persists even after equality is relevant for even lower masses, not shown in this figure.

More generally, the power spectrum is affected indirectly through the oscillation time $a_m$. For fixed $m_a$ and $f_a$, $a_m$ increases—i.e., the onset of oscillations is delayed. This shifts the characteristic momentum $k_\star$ to smaller values, as discussed in \Cref{section:Cut-off}. Because the power spectrum scales as $P_\phi \propto k_\star^{-3}$, the amplitude increases, leading to mildly stronger constraints for a fixed relic density.

\section{Discussion}
\label{sec:discussion}
In this paper, we have developed numerical and analytical techniques to derive constraints on global cosmic strings through searching for their relics in the form of pseudo Nambu-Goldstone bosons. The formalism builds on existing work that treats the scalar field as a summation of plane waves and expands it to treat the case of an expanding universe. In this framework, we derived the power spectrum explicitly and identified the associated transfer function that connects the super-horizon and the sub-horizon modes when the string network collapses. The general form of the former was given in \Eq{eqn:general power spectrum}, while the latter is plotted in \Fig{fig:transfer functions}. We exhaustively investigated the properties of the transfer function and how that depends on the parameters of the systems. The form of the spectrum depends very sensitively on the characteristic momentum of the particles $k_\star$, which is identified with the infrared cutoff of the theory, the inverse size of the co-moving horizon at the time of the collapse of the string network.

In our analysis, we have derived the form of the transfer functions for all scales, while also recovering the white noise plateau for wavenumbers below $k_\star$. We have done this numerically using a time evolving spectral index $q$ \cite{Gorghetto_2018} and we have also done this analytically assuming that the energy spectrum $\Omega_{\phi}(k)$ is independent of $k$, with the result given in \cref{eqn:transfer small k,eqn:transfer large k}. 

Using the likelihood method and power spectrum data from five different experiments, we placed constraints in the parameter space of $m_{a}-f_{a}$ (\Fig{fig:constraints cosmic strings} and \Fig{fig:constraints cosmic strings n=1}), where $f_{a}$ is the energy scale of the U(1) symmetry whose spontaneous breaking gave rise to the pseudo Nambu-Goldstone boson with mass $m_{a}$. Compared to previous works that imposed a cut-off of the spectrum at $k = k_\star$~\cite{Gorghetto:2021fsn,Gorghetto:2025uls}, the constrained parameter space is now significantly expanded for a given observable.

We have also investigated the effects of  introducing the temperature dependence of the boson's mass. The main consequence is that the characteristic momentum $k_\star$ is shifted to lower values for fixed mass. As a result, there is enhanced power for a given relic density and we are able to probe smaller values of $f_{a}$ with a given cosmological dataset. As we have pointed out, our analysis does not apply in the parameter space where the mass has not reached its temperature independent value before radiation-matter equality. One interesting research question would be to investigate how the density perturbations of this non-relativistic component evolve when its mass is still temperature dependent and to reliably extend the constraints of \Fig{fig:constraints cosmic strings n=1} to masses $10^{-25} \textrm{eV} \lesssim m_{a} \lesssim 10^{-23} \rm eV$.     

For smaller masses, $ m_{a} \lesssim 10^{-26} \textrm{eV} $ in \Fig{fig:constraints cosmic strings} the particles remain relativistic even though the string network has collapsed by the time of matter-radiation equality. In this regime there is a substantial dark radiation component present during matter domination. We leave the study of how this dark radiation would influence cosmological observables to future work. 
  
For even lower masses, $m_a \lesssim 10^{-28} {\rm eV}$, the string network persists after equality, providing a further modification to the CMB anisotropies, an observation that has led to the constraint of the string tension $G \mu_{\rm eff} \lesssim 10^{-7}$, where $G$ is Newton's constant and $\mu_{\rm eff}$ is the effective tension of cosmic strings~\cite{Charnock:2016nzm,Lizarraga:2016onn,Lopez-Eiguren:2017dmc,Pen:1997ae,Pogosian:1999np}. In particular, the fluctuations induced by cosmic strings are starkly different from the adiabatic fluctuations produced during inflation. While the latter exhibit the well-known acoustic peaks, the former are incoherent and the resulting spectrum features a single bump. It would be interesting to investigate if the constraints we have derived here are competitive to those derived from the anisotropies of the $\rm CMB$ due to the presence of the strings during recombination.

\section{Acknowledgments} 
AK is supported by the Onassis Foundation - Scholarship
ID: F ZS 031-1/2022-2023. The research of JD is supported in part by the U.S. Department of Energy grant number DESC0025569.
\bibliographystyle{apsrev4-2}
\bibliography{sorsamp}
\appendix

\section{Density correlations in expanding universe}
\label{sec:density correl}
In the main text, we derived the density-density correlation functions by building them up from the ultralight dark matter field. To this end, we required the two-point correlation functions of the scalar field $\phi$. These are given by:
\begin{widetext}
    \begin{align}
       & \langle \phi(x) \phi(x') \rangle  = \int \frac{d^{3}k}{(2 \pi)^{3}} F_{k} \cos \left[ \int^{t}_{t'} \omega_{\textbf{k}}(t'') dt'' - \textbf{k} \cdot \Delta \textbf{x} \right] \\  &
        \langle \dot{\phi}(x) \phi(x') \rangle = - \int \frac{d^{3}k}{(2 \pi)^{3}} F_{k} \omega_{\textbf{k}}(t) \sin \left[ \int^{t}_{t'} \omega_{\textbf{k}}(t'') dt'' - \textbf{k} \cdot \Delta \textbf{x} \right] \\  &
        \langle \phi(x) \nabla \phi(x') \rangle = - \int \frac{d^{3}k}{(2 \pi)^{3}} F_{k} \textbf{k} \sin \left[ \int^{t}_{t'}  \omega_{\textbf{k}}(t'') dt'' - \textbf{k} \cdot \Delta \textbf{x} \right]  \\  &
        \langle \dot{\phi}(x) \dot{\phi}(x') \rangle = \int \frac{d^{3}k}{(2 \pi)^{3}} F_{k} \omega_{\bf k}(t) \omega_{\textbf{k}}(t') \cos \left[ \int^{t}_{t'} \omega_{\textbf{k}}(t'') dt'' - \textbf{k} \cdot \Delta \textbf{x} \right] \\ &
           \langle \dot{\phi}(x) \nabla\phi(x') \rangle = -\int \frac{d^{3}k}{(2 \pi)^{3}} F_{k} \omega_{\textbf{k}}(t) \textbf{k} \cos \left[ \int^{t}_{t'} \omega_{k}(t'') dt'' - \textbf{k} \cdot \Delta \textbf{x} \right] \\ &
           \langle \partial_{i}\phi(x) \partial_{j}\phi(x') \rangle = \int \frac{d^{3}k}{(2 \pi)^{3}} F_{k} k_{i} k_{j} \cos \left[ \int^{t}_{t'} \omega_{\textbf{k}}(t'') dt'' - \textbf{k} \cdot \Delta \textbf{x} \right] 
    \end{align}
\end{widetext}
where $F_{k}$ is defined as:

\begin{equation}
    F_k(t,t') \equiv \frac{f(k)}{\sqrt{a^{3}(t) \omega_{\textbf{k}}(t) a^{3}(t') \omega_{\textbf{k}}(t')}}
\end{equation}

It is straightforward to use these field correlators to find the density correlations $\langle \rho(x) \rho(x') \rangle$. As discussed, these split into terms $\langle \rho(x) \rho(x') \rangle = \langle \rho(x) \rho(x') \rangle_{+} + \langle \rho(x) \rho(x') \rangle_{-}$, which have the general form:
\begin{align}
\label{eqn:correlations density}
\begin{split}
    &\langle \rho(x) \rho(x') \rangle_{\pm} = \frac{1}{4} \int \frac{d^{3}k d^{3}k'}{(2 \pi)^{6}}  F_{k} F_{k'} A^{\pm}_{\bf k k'} \times  \\ & \times \cos \left[ \int^{t}_{t'} dt'' (\omega_{\bf k}(t'') \pm \omega_{\bf k'}(t'')) - (\textbf{k} \pm \textbf{k}') \cdot \Delta\textbf{x} \right]
\end{split}
\end{align}
with the coefficients $A^{\pm}_{\bf k k'}$ given explicitly by: 
\begin{widetext}
    \begin{align}
    \begin{split}
        A^{\pm}_{\bf k k'}  = 
         & m^{2}(t) m^{2}(t')+\omega_{\bf k}(t) \omega_{\bf k}(t') \omega_{\bf k'}(t) \omega_{\bf k'}(t') \mp  m^{2}(t')\omega_{\bf k}(t) \omega_{\bf k'}(t) \mp m^{2}(t) \omega_{\bf k}(t') \omega_{\bf k'}(t') +  \\ & + \textbf{k} \cdot \textbf{k}' \left( \frac{\omega_{\textbf{k}}(t) \omega_{\textbf{k}'}(t) \mp m^{2}(t)}{a^{2}(t')} + \frac{\omega_{\textbf{k}}(t') \omega_{\textbf{k}'}(t') \mp m^{2}(t')}{a^{2}(t)} \right) + \frac{\left( \textbf{k} \cdot \textbf{k}'\right)^{2}}{a^{2}(t) a^{2}(t')}
         \end{split}
    \end{align}
\end{widetext}
In the non-relativistic limit where $\omega_{\bf k}(t) \simeq m(t) + \textbf{k}^{2}/2 m a^{2}(t)$, these coefficients become $A^{+}_{\bf k k'} \simeq \left( |\textbf{k} + \textbf{k}'|^{2}  \right)^{2}/4 a^{2}(t) a^{2}(t')$ and $A^{-}_{\bf k k'} \simeq 4 m^{2}(t) m^{2}(t')$.

\section{The Maxwell-Boltzmann distribution}
\label{sec:maxwell}
It is illuminating to compare the cosmic string spectrum to that of the familiar Maxwell-Boltzmann distribution. To that end, we briefly summarize the results in this case. We assume that the particles start to behave non-relativistically at time $t_{m}$, with corresponding scale factor $a_{m}$, and that their phase space distribution evolves self-similarly after that. We evaluate \Eq{eqn:correlations density} at $t=t'=t_{m}$.  We may first write the integrals and the phase space densities in terms of the velocity $v$:
\begin{align}
\begin{split}
   & f\left( k \right)  \frac{d^{3} k}{a^{3}_{m} (2\pi)^{3}}=  \frac{\bar{\rho}(t_{m})}{m} f\left( v \right) d^{3} v.
\end{split}
\end{align}
We have defined $f(v)$ in such a way that $\int d^{3}v f(v) = 1$.
Plugging this in \Eq{eqn:correlations density} and focusing on the slow mode, we get:
\begin{align}
\begin{split}
     &\langle \rho_{\phi}(\textbf{x}) \rho_{\phi}(\textbf{x}') \rangle_{-} \\&=  \bar{\rho}^{2}(t_{m})  \int   d^{3}v d^{3}v'  f(v) f(v') 
    \cos \left( m\Delta \textbf{x} \cdot \Delta \bf{v} \right),
\end{split}
\label{eqn:rho correl approx}\end{align}
where we defined $\Delta \textbf{v} = \textbf{v} - \textbf{v}'$.

The Maxwell-Boltzmann distribution is given by $f(v) =  e^{-v^{2}/2 v^{2}_{0} a^{2}_{m}}/(2 \pi v^{2}_{0} a^{2}_{m})^{3/2}$. Plugging this in \Eq{eqn:rho correl approx} and employing the transformations $\textbf{v}_{\pm} = (\textbf{v} \pm \textbf{v}')/\sqrt{2}$ to simplify the integrals, we get: 

\begin{align}
\begin{split}
     \langle \rho_{\phi}(\textbf{x}) \rho_{\phi}(\textbf{x}') \rangle_{-} =  & \frac{\bar{\rho}^{2}_{\phi}(t_{m})}{\left( 2\pi v^{2}_{0} a^{2}_{m} \right)^{3}}  \int d^{3}v_{+} d^{3}v_{-} e^{- \frac{v_{+}^{2} + v_{-}^{2}}{2 v^{2}_{0} a^{2}_{m}}} 
   \\ & 
    \cos \left( \sqrt{2} \Delta \textbf{x} \cdot \textbf{v}_{-} \right),
\end{split}
\label{eqn:rho correl v+}
\end{align}
These integrals can be done analytically and we obtain:
\begin{equation}
     \langle \rho_{\phi}(\textbf{x}) \rho_{\phi}(\textbf{x}') \rangle_{-} =  \bar{\rho}_{\phi}^{2}(t_{m}) e^{  - k^{2}_\star \Delta \textbf{x}^{2}},
\label{eqn:density correl space}
\end{equation}
where $k_\star = a_{m} m v_{0}$.
The power spectrum is then:
\begin{align}
\begin{split}
    P_{\phi}(k,t_{m}) \equiv  &  \int d^{3} x e^{-i \textbf{k} \cdot \textbf{x}}  \frac{\langle \rho_{\phi}(\textbf{x}) \rho_{\phi}(0) \rangle}{\bar{\rho}^{2}_{\rm DM}(t_{m})}  = \\ & = \left( \frac{\bar{\rho}_{\phi}(t_{0})}{\bar{\rho}_{\rm DM}(t_{0})} \right)^{2} \pi^{3/2} \left( \frac{1}{k_\star} \right)^{3} \mathcal{T}_{\rm MB} \left( \frac{k}{k_\star}\right),
\end{split}
\label{eqn:uldm power spectrum}
\end{align}
where we considered that the dark matter as well as the ultra light axion's density both scale as $a^{-3}$ and defined the transfer function $\mathcal{T}_{\rm MB}(y) = e^{-y^{2}/4}$. This is the transfer function that we depict in \Fig{fig:transfer functions}. 




\section{The scaling regime}
\label{sec:scaling regime}
In this appendix, we review the essentials for the calculation of the spectrum of axions emitted by cosmic strings. The discussion follows closely Ref.~\cite{Gorghetto_2018}. 

Before the axions acquire mass, at a time that is determined by the solution of $H(a_{ m}) = m$, they are emitted by strings, whose energy density is given by \cite{Gorghetto_2018,Sikivie_2008}:
\begin{equation}\label{eqn: 5.1}
    \rho_{s} = \frac{\xi(t) \mu_{\textrm{eff}}}{t^{2}} 
\end{equation}
where $\xi(t)$ is the average number of strings per Hubble volume and $\mu_{\rm eff}$ is the effective string tension which we will write shortly. In the scaling regime, $\xi$ tends to a particular value which is insensitive to the initial conditions and is usually parameterized by:
\begin{equation}\label{eqn: 5.2}
    \xi(t) = \alpha \text{log} \left( \frac{m_{\textrm{r}}}{H} \right) + \beta
\end{equation}
where $\alpha$ and $\beta$ are constants and $m_{\textrm{r}}$ is the inverse core size of the string, which we will assume to be $m_{r} \sim f_{a}$. The effective tension is given by:
\begin{equation}\label{eqn: 5.3}
    \mu_{\textrm{eff}} (t) = \pi f^{2}_{a} \log \left( \frac{m_{r} \gamma}{H \sqrt{\xi}} \right)
\end{equation}
where $\gamma$ is of order unity. The rate of emission of axions from strings in the large log limit is given by \cite{Gorghetto_2018,Dror_2021}:
\begin{equation}\label{eqn: 5.4}
    \Gamma_{\phi} \rightarrow 2 H \frac{\rho_{s}}{\rho_{\textrm{SM}}} \rho_{\textrm{SM}}
\end{equation}
where $\rho_{\rm SM} = \frac{\pi^{2}}{30} g_{\ast}(T) T^{4}$ is the energy density in the Standard Model and $g_{\ast}(T)$ are the relativistic degrees of freedom. Using \Eq{eqn: 5.1} and the relation $t = \frac{1}{2H}$, we find for the ratio of the string density to the standard model density: 
\begin{equation}\label{eqn: 5.5}
    \frac{\rho_{s}}{\rho_{\rm SM}} = \frac{4 \xi \mu_{\rm eff}}{3 M^{2}_{\rm pl}}
\end{equation}
We can write the density and the spectrum of these axions as $\rho_{\phi} = \int dk  \frac{\partial \rho_{\phi}}{\partial k}$. The spectrum can also be written as:
\begin{equation}\label{eqn: 5.6}
    \frac{\partial \rho_{\phi}}{\partial k} = \int^{t_{m}}_{t_{\rm min}} dt' \frac{\Gamma_{\phi}(t')}{H(t')} \frac{a^{3}(t')}{a^{4}(t)}  F\left[ \frac{k'}{H'} , \frac{m_{r}}{H'} \right]
\end{equation}
where $k' =  \frac{k}{a(t')}$ and $t_{\rm min}$ is the start of the scaling regime.~\footnote{Note that in our notation $k$ is a co-moving momentum,  while in \cite{Gorghetto_2018} it denotes the physical momentum, and hence the absence of the scale factor $a(t)$ in our definition of $k'$.} Defining $x=\frac{k}{a H}$ and $y = \frac{m_{r}}{H}$, we can parameterize the spectrum as:
\begin{align}\label{eqn: 5.7}
    \begin{split}
    F[x,y] & = \frac{\mathcal{N}}{x^{q}}, \hspace{0.3cm} x_{_{\rm IR}} < x < y \\ &
    = 0, \hspace{0.3cm} \text{otherwise}.
    \end{split}
\end{align}
$x_{_{\rm IR}}$ is an infrared cut-off of the order $x_{\rm IR} \simeq 10$, $y = \frac{m_{r}}{H}$ is the ultra-violet cut-off and $\mathcal{N}$ is a constant that ensures the spectrum is normalized according to $\int dx F[x,y] = 1$. When $q>1$, the spectrum is IR dominated while when $q<1$ it is UV dominated. The form of the spectral index that we use is \cite{Gorghetto:2020qws}:

\begin{equation}
\label{eqn:q}
    q(a) = 0.51 + 0.053 \log \left( \frac{f_{a}}{H(a)} \right).
\end{equation}

Using the definition \Eq{eqn:Omega def}, \Eq{eqn: 5.4}, \Eq{eqn: 5.5} and the definition of the Hubble constant to express the integral in terms of the scale factor, we get:
\begin{equation}
    \Omega_{\phi} (k,a) = \frac{8 k}{3 a^{4} M^{2}_{\rm pl} \rho_{c}} \int^{a}_{a_{\rm min}} da' \frac{(a')^{2} \xi' \mu'_{\rm eff} \rho'_{\rm SM}}{H'} F\left[ \frac{k'}{H'}, \frac{f_{a}}{H'} \right].
\label{eqn:Omega}\end{equation}
where $a_{\rm min}$ is the beginning of the scaling regime.
Integrating over the wavenumbers $k$, we obtain the relativistic energy density at time $a$:

\begin{equation}
    \rho_{\phi}(a) = \frac{8}{3 M^{2}_{\rm pl} a^{4}} \int^{a}_{a_{\rm min}} da' (a')^{3} \xi' \mu'_{\rm eff} \rho'_{\rm SM}.
\label{eqn:relativistic energy density}\end{equation}

\section{Infrared cut-off of spectrum}
\label{section:Cut-off}

The infrared cutoff that was given in \cref{eqn:kir}, may depend on $f_{a}$ if $n \neq 0$. Parametrically, we have for $a_{m}$:

\begin{equation}
    a_{m} \propto \frac{a_{\rm eq} T_{\rm eq}}{\sqrt{m_{a,0} M_{\rm pl}}} \left( \frac{M_{\rm pl}}{f_{a}} \right)^{\frac{n}{2n+4}}
\end{equation}

where we used the definition of $a_{m}$ through $H(a_{m}) = m(a_{m})$ and:

\begin{equation}
    H(a_{m}) \propto \frac{T^{2}_{\rm eq} a^{2}_{\rm eq}}{M_{\rm pl} a^{2}_{m}}
\end{equation}

where $T_{\rm eq} \simeq 0.1 \rm eV$ is the temperature at the time of the radiation-matter equality. The infrared cutoff then scales as:
\begin{equation}
k_{\rm IR} \propto a_{\rm eq} T_{\rm eq} \sqrt{\frac{m_{a,0}}{M_{\rm pl}}} \left( \frac{f_{a}}{M_{\rm pl}}\right)^{\frac{n}{2n+4}} x_{\rm IR}
\end{equation}

Therefore, for the same mass and $x_{\rm IR}$, the effect of temperature dependence is to shift the infrared cut-off to smaller energies, in the physically relevant regime $f_{a} < M_{\rm pl}$. 

\section{Free-streaming}
\label{sec:free streaming}
To determine the free-streaming constraint, we follow \cite{amin2024lowerbounddarkmatter,Liu:2024pjg,Long:2024cak}. The free streaming length is given by \Eq{eqn:free streaming scale}. 
This integral can be determined analytically for any scale factor and momentum $k$, in the case where the mass is temperature independent, and the result is (see appendix A of \cite{Liu:2024pjg}):

\begin{equation}
    \lambda_{\rm fs}(k,a) = \frac{\sqrt{2}}{H_{\rm eq} a_{\rm eq}} F\left(\hat{k},y\right)
\end{equation}

where $\hat{k} \equiv k/m a_{\rm eq}$, $y \equiv a/a_{\rm eq}$ and $F(\hat{k},y)$ is a function that is given in the appendix A of \cite{Liu:2024pjg}. For non-relativistic momenta at equality, $\hat{k} \ll 1$, the dominant contribution to the free-streaming length comes from $y \ll 1$ and it is logarithmic in $y$. For temperature-dependent masses, the integral can be performed numerically.

As it was shown in \cite{amin2024lowerbounddarkmatter}, free streaming affects only the adiabatic part of the power spectrum and it is encoded in the transfer function:

\begin{equation}
\label{eqn:transfer fs}
    T_{\rm fs}(k,a) = \frac{m^{2}_{\rm NR}}{\bar{\rho}_{\rm DM}(a)}\int \frac{d^{3} p}{(2 \pi)^{3}}\mathcal{P}(p) \sinc \left( p \lambda(p,a) \right)
\end{equation}

where $\mathcal{P}(p)$ is an initial power spectrum of fluctuations. This is related to the usual phase space density through \cite{Long:2024cak}:

\begin{equation}
\label{eqn:P to f}
    f(p,t) = a^{3}(t) \omega_{\textbf{p}}(t) \mathcal{P}(p)
\end{equation}

Relating this phase space density to our spectrum $\Omega_{\phi}$ through \Eq{eqn:f and Omega} and plugging it in \Eq{eqn:transfer fs}, we obtain:

\begin{equation}
    T_{\rm fs}(k,a) = \frac{m_{\rm NR} \rho_{c}(a_{\rm NR}) a^{4}_{\rm NR}}{\bar{\rho}_{\rm DM}(a) } \int \frac{dp}{p^{2}} \Omega_{\phi}(p) \sinc \left( k \lambda(p,a) \right)
\end{equation}

In the case where the spectrum $\Omega_{\phi}(k)$ is dominated  at $k_\star$, the integral can be simplified to give \Eq{eqn:transfer free approx} where we also used \Eq{eqn:relic density}.
To draw the bound, we performed the integral in \Eq{eqn:transfer fs} numerically, evaluated at the present time, and used the spectrum given in \Eq{eqn:Omega}.

\end{document}